\documentclass{sigchi}

\CopyrightYear{2020}
\setcopyright{acmlicensed}
\doi{https://doi.org/10.1145/3313831.3376244}
\isbn{978-1-4503-6708-0/20/04}
\conferenceinfo{CHI'20,}{April  25--30, 2020, Honolulu, HI, USA}
\acmPrice{\$15.00}

\usepackage{balance}       
\usepackage{graphics}      
\usepackage[T1]{fontenc}   
\usepackage{txfonts}
\usepackage{mathptmx}
\usepackage[pdflang={en-US},pdftex]{hyperref}
\usepackage{color}
\usepackage{booktabs}
\usepackage{textcomp}

\usepackage{microtype}        
\usepackage{ccicons}          

\usepackage{todonotes}
\usepackage{enumitem}

\newcommand*{\affmark}[1][*]{\textsuperscript{#1}}

\def\plaintitle{AutoGain: Gain Function Adaptation with Submovement Efficiency Optimization}

\def\emptyauthor{}
\def\plainkeywords{Pointing; Submovement; CD gain functions; Pointer
acceleration; Human Performance; Pointing facilitation.}

\makeatletter
\def\url@leostyle{%
  \@ifundefined{selectfont}{
    \def\UrlFont{\sf}
  }{
    \def\UrlFont{\small\bf\ttfamily}
  }}
\makeatother
\def\pprw{8.5in}
\def\pprh{11in}

\definecolor{linkColor}{RGB}{6,125,233}
\hypersetup{%
  pdftitle={\plaintitle},
  pdfauthor={\emptyauthor},
  pdfkeywords={\plainkeywords},
  pdfdisplaydoctitle=true, 
  bookmarksnumbered,
  pdfstartview={FitH},
  colorlinks,
  citecolor=black,
  filecolor=black,
  linkcolor=black,
  urlcolor=linkColor,
  breaklinks=true,
  hypertexnames=false
}

\definecolor{bcolor}{rgb}     {0.2,0.8,0.8}
\definecolor{mcolor}{rgb}     {0.0,0.5,0.4}
\definecolor{acolor}{rgb}     {0.0,0.3,0.9}

\newcommand\bl[1]{\textcolor{bcolor}{[B: #1]}}

\newcommand\mn[1]{\textcolor{mcolor}{[M: #1]}}

\newcommand\ao[1]{\textcolor{acolor}{[A: #1]}}

\definecolor{fixcolor}{rgb}	{.7, .0, .0}

\renewcommand{\todo}[1]{}
\renewcommand{\mn}[1]{}
\renewcommand{\bl}[1]{}
\renewcommand{\ao}[1]{}
\usepackage{eurosym}

\newcommand{\finalAdd}[1]{#1}
\newcommand{\finalDel}[1]{}

\newcommand\etal[0]{{\textit et al.}}

\definecolor{Gray}{gray}{0.9}

\newcommand\block[0]{{\sc Block}}
\newcommand\refpart[0]{{\sc Reference}}
\newcommand\golfpart[0]{{\sc \methodname}}

\newcommand\sigpart[0]{{\sc Sigmoid}}

\newcommand{\methodname}{AutoGain}

\begin{document}

\title{AutoGain: Gain Function Adaptation with Submovement Efficiency Optimization}


\numberofauthors{1}
\author{%
  \alignauthor{Byungjoo Lee\affmark[1,2], Mathieu Nancel\affmark[3,2], Sunjun Kim\affmark[1,2], and Antti Oulasvirta\affmark[2]\\
    \affaddr{\affmark[1]KAIST, \affmark[2]Aalto University, \affmark[3] Inria \& Univ. Lille, UMR 9189 - CRIStAL, Lille, France}\\    
    \email{byungjoo.lee@kaist.ac.kr, mathieu.nancel@inria.fr, \{sunjun.kim,antti.oulasvirta\}@aalto.fi}}\\
}

\teaser{
  \centering
  \includegraphics[width=1.0\textwidth]{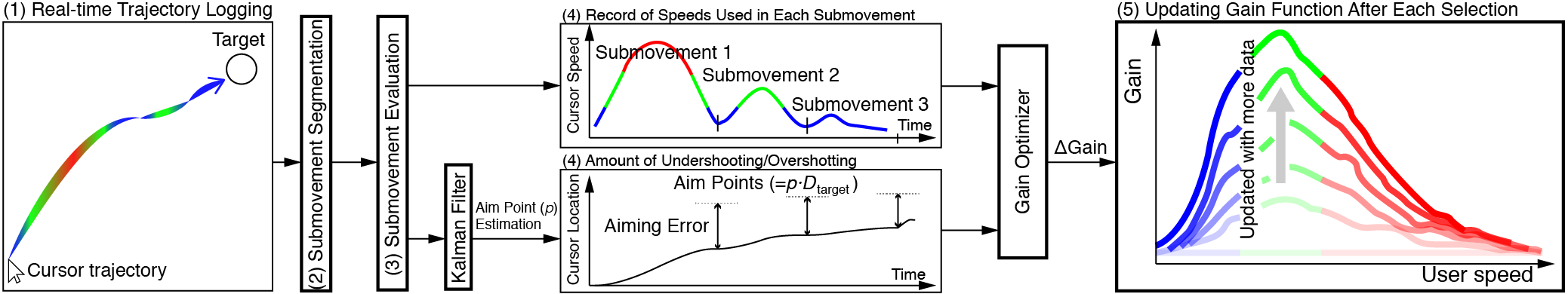}
  \caption{
      \methodname~ adapts transfer function on indirect pointing devices by using a novel submovement-level tracking-and-optimization approach. 
      It gradually updates the control-to-display (CD) gain function based on analysis of speed and error in a user's submovements. 
      It updates the CD gain function trying to minimize expected aiming error for typical submovements.}
  \label{fig:moneyshot}
}
\maketitle

\begin{abstract}
A well-designed control-to-display gain function can improve pointing performance with indirect pointing devices like trackpads. 
However, the design of gain functions is challenging and mostly based on trial and error. 
\methodname~is a novel method to individualize a gain function for indirect pointing devices in contexts where cursor trajectories can be tracked. 
It gradually improves pointing efficiency by using a novel submovement-level tracking+optimization technique that minimizes aiming error (undershooting/overshooting) for each submovement.
We first show that \methodname~can produce, from scratch, gain functions with performance comparable to commercial designs, in less than a half-hour of active use. 
Second, we demonstrate \methodname's applicability to emerging input devices (here, a Leap Motion controller) with no reference gain functions.
Third, a one-month longitudinal study of normal computer use with \methodname\ showed performance improvements from participants' default functions.
\end{abstract}




\section{Introduction}

\emph{Control-to-display (CD) function} is a pointing facilitation technique used in the vast majority of indirect input devices like mice and trackpads.
A CD function determines the speed of a control point in display space (e.g. a cursor) from the user's motion speed in control space (e.g. a mouse speed).
A gain function ($f_{C\!D}$) computes a scalar factor (or ``gain'') from the instantaneous input speed ($v_{in}$), that is multiplied to that input speed to obtain the speed of the control point ($v_{out}$):
$v_{out}=f_{C\!D}(v_{in})\times v_{in}$.
The type and design of a gain function can affect the kinematics  \cite{lee2013kinematic,lee2015mouse} and performance \cite{casiez2011no,casiez08,nancel15,anglemouse} of pointing.
\emph{Speed-dependent} CD gain functions are interesting to performance-oriented users, such as gamers and information workers.
They affect which limbs are used \cite{accot2001scale, zhai1996influence},
reduce unnecessary clutching \cite{jellinek1990powermice}.
and allow users to avoid constraining high movement speeds that otherwise might lead to high error \cite{casiez08}.
Some studies show improvements in selection time of up to 24\% with speed-dependent gain functions \cite{casiez2011no}, 
compared to constant gains.

Prior to this work, speed-dependent gain functions have been designed based on either trial and error \cite{casiez2011no} or heuristic iteration \cite{nancel15,yun2015interactivity} (see Related Work).
While methods exist for constant functions~\cite{casiez08}, no automated or computer-assisted method exists for the general, speed-dependent case.
Calibration-free methods exist, e.g. \cite{prism}, but they are known to not scale well \cite{nancel13}.
One challenge is that the design space is large: 
in principle, any continuous function is valid.
A function must strike a balance between high and low gains that control the trade-off between the speed and accuracy of pointing.
A high CD gain reduces the time to approach a distant target, but it may hamper precise positioning on top of a smaller target.
Low gains increase precision but possibly at the cost of slower approaches.
One also has to solve how to adapt the function:
How to infer data from a user and set the right speed while ensuring stable performance given that users also learn?

This paper investigates a novel computational approach to adapting control-to-display (CD) gain functions to individuals. 
\emph{\methodname} addresses the two challenges  (Figure~\ref{fig:moneyshot}).
It iteratively adapts the CD gain function after each target acquisition trial, using the estimated accuracy of its submovements.
The submovement optimization approach is informed by a theory of motor control. 
It tries to minimize aiming error (overshooting/undershooting). This is calculated with respect to the user's inferred aim point for each submovement within the last target acquisition (see bottom of Figure \ref{fig:moneyshot}).
The gain function is discretized, with input speeds ``binned'' like in a histogram, and the gain of each bin is modified based on the speeds used in the previous pointing act.
Unlike previous methods, the method requires no human supervision and minimal initialization: after setting the overall speed of convergence, it starts from 
any function and adapts it based on automated observations and principles.
This makes it deployable for devices for which no manually designed or reference functions exist.
%

 To sum up, 
\methodname\ is the first method to automatically optimize a gain function from a user's actual motion data, using minimal assumptions about the shape of that function.
In the rest of the paper, we detail the method, and report the results of three evaluative studies. 

\section{Related Work}

\finalAdd{Despite extended early work on force-to-motion functions for isometric input (e.g. \cite{rutledge1990force}), and although virtually every study using an isotonic} indirect pointing device uses \finalAdd{some form of speed-based} transfer function,
the design and adaptation of the latter remains a relatively little-studied topic \cite{casiez2011no,nancel15,yun2015interactivity}.

Their simplest form is fixed gains, i.e. constant ratios between input and output movement velocities.
Casiez et al. \cite{casiez08} presented a principle to choose usable ranges of fixed gain values.
A minimum gain $C\!D_{min}$ should allow the user to acquire the most distant targets without clutching,
and a maximum gain $C\!D_{max}$ should allow accessing each individual pixel.
Nancel et al. later proposed a formulation that relaxes these constraints \cite{nancel2015mid}, allowing for some clutching and arbitrarily small targets.

%
%
%
%
%
Most commercial gain functions for mice and trackpads are not constant, but little research has been conducted to guide the design of such (speed-dependent) gain functions.
It is generally agreed that gain functions can perform better than constant gains, even though that is usually derived from comparing one gain value against one or more gain functions, e.g.~\cite{casiez2011no,adaptive}.
In effect, systematic comparisons are scarce.
Casiez and colleagues~\cite{casiez08} compared 6 gain values and 6 gain function settings, and report a borderline-significant effect ($p=.065$) with an improvement of 3.3\% in selection time between the best settings of fixed vs. non-constant gains.

Casiez and Roussel \cite{casiez2011no} reverse-engineered gain functions in existing operating systems.
They found that all commercially deployed functions share some features, namely monotonic increase in the beginning and comparable maximum outputs.
They also found differences, especially in minima, continuity, and shape of the functions. 

Nancel and colleagues~\cite{nancel13,nancel15} proposed a generic gain function based on the generalized logistic curve.
It expresses four features of gain functions: 
the asymptotic minimum and maximum output gains, the abscissa (input velocity) of the curve's inflexion point, and its slope at that inflexion point. 
This function was successfully applied to translation- and rotation-based input channels, and later on to different input devices and interactive environments~\cite{haque15,liu15}.
However, tuning the parameters of the function is ad hoc:
initial values are based on heuristics derived from~\cite{casiez08}, and suitability is left to designer's judgment.
%
%
Earlier approaches combined simple sine-based CD gain functions with absolute position control \cite{prism,adaptive,smoothed}, with an implicit velocity-based transition.
%

%

In contrast to previous studies considering how the magnitude of input movement is being transferred, several studies have investigated other aspects of control-to-display relationships, such as angular deviation~\cite{anglemouse}, coordinate disturbance \cite{lee2015mouse}, proximity to target~\cite{semantic} and movement direction \cite{lee2013kinematic}. 
Results include accurate and faster drawing \cite{lee2015mouse}, reduced overshoot \cite{lee2013kinematic}, and better experience for motor impaired users \cite{anglemouse}.

To sum up, speed-dependent gain functions (a.k.a. acceleration) are expected to be superior to constant gains.
However, the design of successful gain functions for any setup has been limited to  hand-tuned approaches.
Research suggests that there is value in analyzing pointing kinematics in their design.

\section{Overview of Technique}

\begin{figure}[h]
  \centering
  \includegraphics[width=0.95\columnwidth]{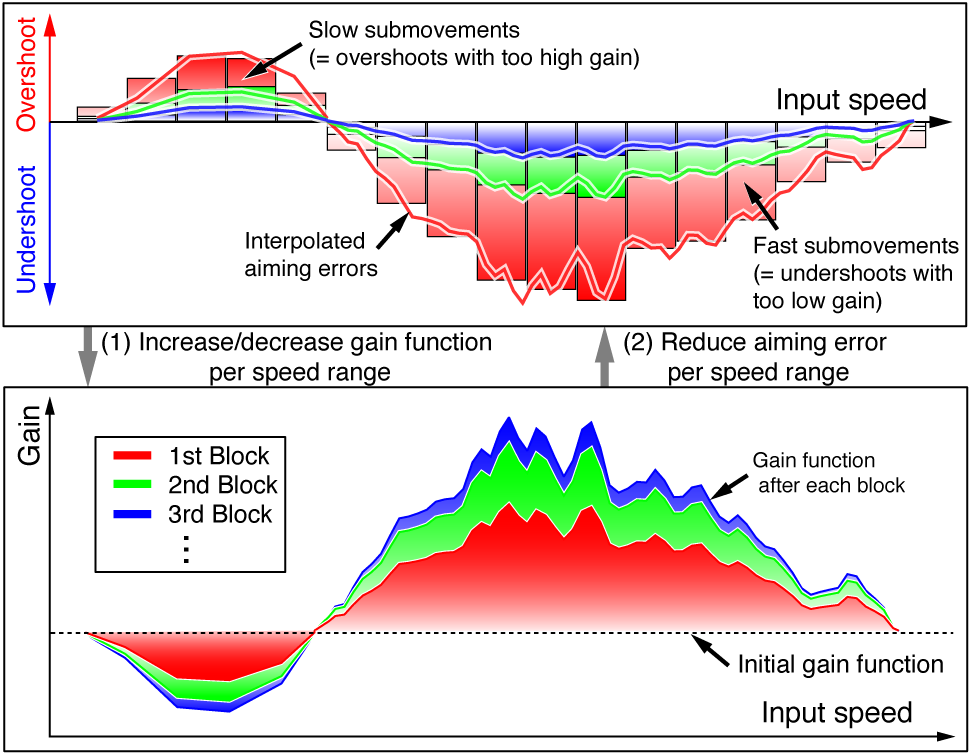}
  \caption{\methodname\ improves a discrete gain function by updating gains locally per speed range.
  It segments a movement into submovements to update a profile of 
  aiming errors, which it tries to reduce by adapting the gain response. 
  Here, it is shown how \methodname\ fixes too-low gains at high input speeds, and too-high gains at low input speeds. 
  }
\label{fig:principle}
\end{figure}

\methodname\ is designed under the assumption that \emph{the aiming error of individual pointing submovements is an indication of the inadequacy of the gain function}:
if a submovement ends before its intended aim point (undershoot), then the gain function was too low and should be increased; 
if it ends further than its intended aim point (overshoot), then the gain function was too high and should be decreased.
In both cases, the gain change should be proportional to the amount of aiming error.
\methodname\ applies this principle in a speed-dependent manner:
it treats gain functions as series of gains associated to discrete intervals of input speeds, and only alters a gain when the corresponding ``speed bin'' was used in a pointing task.
This procedure ensures it does not change the function erratically.
The principle is summarized in Figure~\ref{fig:principle}.

\methodname\ subscribes to a local optimization scheme that assumes that there exists a case-specific optimum gain function (or \finalDel{several optimal gain functions on a Pareto front} \finalAdd{several optimal gain functions with different trade-offs}) for a user performing pointing tasks of certain difficulty and scale ranges, with a given pointing device.
It also assumes that this function is reachable from an initial constant or speed-dependent gain function, by means of a sequence of local updates to the function.
Moreover, when updates are based on repeated observations of the user, a reasonable estimate of the optimum function can be obtained despite several sources of variability in the human motor system.

To estimate the optimum, \methodname\ builds on two assumptions in earlier theories of human motor control:
\begin{figure}[b]
\centering
\includegraphics[width=1.0\columnwidth]{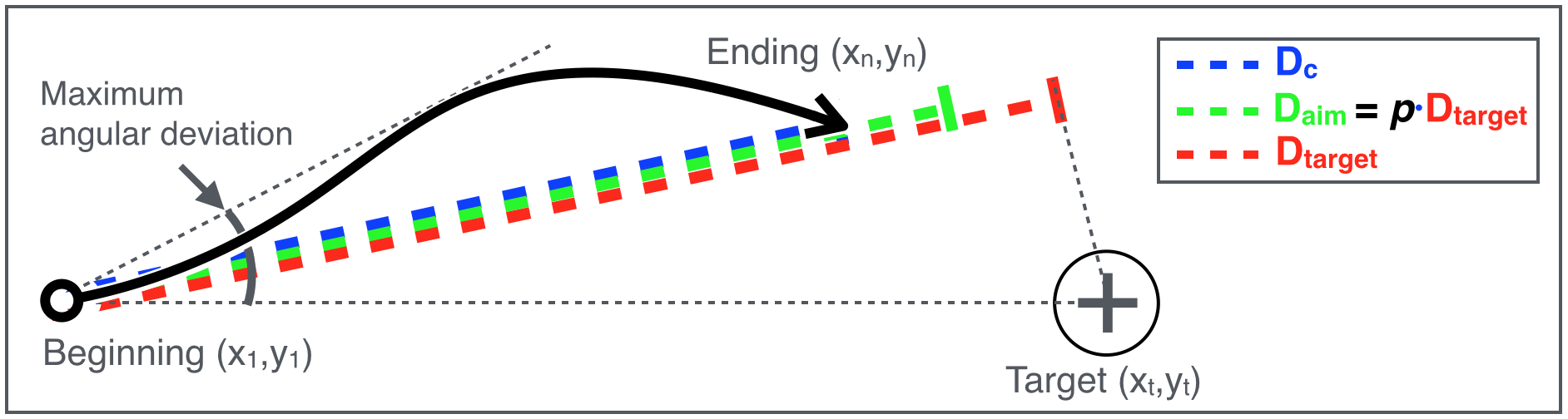}
\caption{\methodname~assumes that each submovement has an implicit aim point ($D_{aim}$) located at a fraction ($p$) to the remaining target distance ($D_{target}$). In this figure, a submovement is undershooting, which is marked with a positive aiming error $R=(D_{aim}-D_{c})$.}
\label{fig:geometry}
\end{figure}
%
%
\begin{itemize}
  \item \textbf{Submovement decomposition}: An aimed movement can be divided into one or more submovements \cite{crossman1983feedback, meyer1988optimality,wisleder2007role} using local accelerations and decelerations in the speed profile.
%
  \item \textbf{Implicit aim point}: A submovement $i$ has an implicit aim point 
	located at a fraction ($p_i$) of the remaining distance to the target ($D_{target,i}$) at its beginning 
	\cite{crossman1983feedback}:
  \begin{equation}
  D_{aim,i}=p_i\cdot D_{target,i}
  \label{eq:p}
  \end{equation}
  Due to stochastic noise in the motor system, submovement endpoints are distributed around the center of the aim point \cite{meyer1988optimality}. With respect to the aim point, the aiming error $R_i$ of a submovement $i$ is defined as: 
  \begin{equation}
  R_i=(D_{aim,i}-D_{c,i})
  \label{eq:r}
  \end{equation}
  With $D_{c,i}$ the distance moved during the submovement $i$ (projected, see Figure~\ref{fig:geometry}). 
  Negative errors indicate overshooting, positive errors indicate undershooting. 
  %
  %
  %
\end{itemize}


These assumptions exploit two well-known theories of aimed movements.
In Crossman and Goodeve's \emph{deterministic iterative corrections model} \cite{crossman1983feedback},
aimed movements are modeled as series of ballistic, open-loop submovements aimed at a constant fraction ($p$) of the remaining distance to the target.
However, because empirical data showed large variations in the duration and aim point of these submovements, an extension of the idea was proposed.
Meyer's \emph{stochastic optimized submovement model} \cite{meyer1988optimality} assumes that neuromotor noise causes the primary submovement to either undershoot or overshoot the target, which requires a corrective submovement to finally reach the target center.
\section{Implementation}

\methodname~updates the gain function after every target selection. 
The updating procedure consists in five steps (see Figure \ref{fig:moneyshot}).
First, it records a pointing trajectory from onset until the target selection. 
Second, it  segments the trajectory into submovements using kinematic criteria. 
Third, it filters submovements out based on their trajectory and dynamic properties. 
Fourth, it computes the amount of aiming error (undershooting/overshooting) for each submovement, relative to its estimated aim point, as well as the input speeds that were used in that submovement.
Fifth, it updates the gain function around these input speeds, using the amplitude and nature of these errors, following a local optimization scheme.

\subsection{Step 1: Real-Time Trajectory Logging} 
\methodname~logs two different time vectors
in real-time, for each input event $t$: (1) the raw input stream $(dx_t,dy_t)$ (in counts) from the input device, and (2) the cursor trajectory $(x_{c,t},y_{c,t})$ (in pixels).
The raw input stream is used to obtain records of movement speeds in submovements. 
The cursor trajectory is used to segment submovements.
We express the relationship between inputs and outputs as follows:
\begin{equation}
  \begin{split}
    v_t&=C_{in}\sqrt{dx^{2}_t+dy^{2}_t} \\
    (x_{c,t+1},y_{c,t+1})&=(x_{c,t},y_{c,t})+C_{out}\cdot C_{in}\cdot G[v_t] \cdot (dx_t,dy_t)\\
  \end{split}
  \label{eq:inputlogging}
\end{equation}
\begin{center}
	  with $C_{in}=\text{Freq}_{in}/\text{Res}_{in}$ and $C_{out}=\text{Freq}_{in}/\text{Res}_{out}$
\end{center}
$C_{in}$ is a factor converting unit of raw input (count) to m/s, which is determined from the input device's resolution (Res$_{in}$, here in points per millimeters) and update frequency (Freq$_{in}$, here in number of events per millisecond) \cite{casiez2011no}.
$C_{out}$ uses the opposite principle to convert transformed input movements in m/s into pixels translations, using the same update frequency Freq$_{in}$ and the resolution of the display (Res$_{out}$, in pixels per millimeter).
\methodname\ treats gain functions as arrays of gains associated to discretized intervals of input speed.
Gains values ($G[v_t]$) are therefore interpolated when $v_t$ is not one of the discretized input values.
 
\subsection{Step 2: Submovement Segmentation} 

\begin{figure}[h]
  \centering
  \includegraphics[width=1.0\columnwidth]{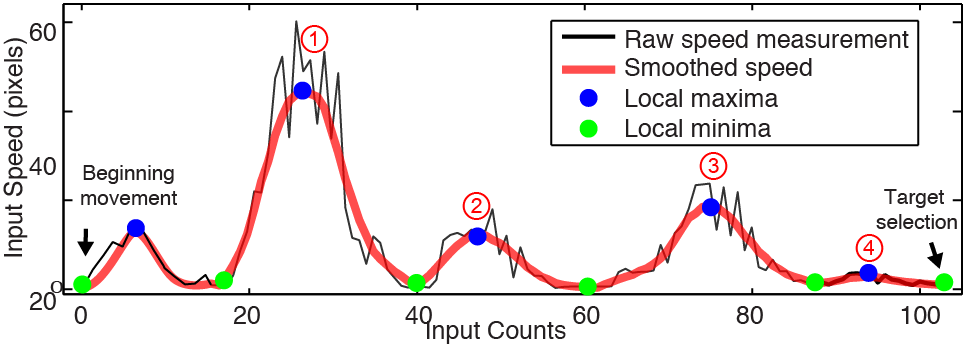}
  \caption{
  	For each trial, \texttt{Persistence1D} identifies submovements as minimum-maximum-minimum triplets in the smoothed input speed profile, then excludes the ones preceding the highest maximum (here \textcircled{1}).
  }
  \label{fig:submovement} 
  ~\vspace{-5mm}
\end{figure}

From the beginning of a movement to the selection of the target, the coordinates are segmented into submovements based on
local extrema in the cursor speed profile~\cite{evans12}, using the \texttt{Persistence1D} \cite{persistence1d} algorithm.
%
%
%
It first smooths the speed profile (red curve in Figure~\ref{fig:submovement}) using a Gaussian kernel filter ($\sigma=3$), then returns all pairs of minima and maxima that exceed a pre-defined persistence value (0.2).
In most cases, resolution of raw inputs are available in integer steps so we set value of persistence less than 1.0 to ensure enough sensitivity on smaller submovements even after the smoothing process.

After identifying the local minima and maxima in the speed profile, each neighboring minimum-maximum-minimum triplet is considered to be a possible submovement. 
As in \cite{evans12}, we only consider submovements from the highest local maximum (\textcircled{1} in Figure~\ref{fig:submovement}), which is assumed to be the initial ballistic movement to the target. 
This helps exclude possible non-aiming movements during the trial.

%
\subsection{Step 3: Submovement Evaluation} 
%
\methodname\ then identifies some unwanted characteristics of the submovements' trajectories: \emph{unaimed}, \emph{interrupted}, and \emph{non-ballistic}.
The goal of this classification is to filter out submovements that are likely to introduce noise in the updates of the gain function, or of the aim point (see Table \ref{tab:taxonomy}).

We define \emph{unaimed} submovements using two tajectory properties: 
(1) the maximum angular deviation, and 
(2) the amount of overshoot (see Figure \ref{fig:geometry}).
Maximum angular deviation is defined as the maximum angle between 
the line joining the first and last points of the submovement's trajectory,
and the line joining the first and any other point of the submovement's trajectory.
The amount of overshoot is defined as: $\max (D_{c} - D_{target}, 0)$.
Any submovement that satisfies at least one of the following conditions is marked as \emph{unaimed}: 
\begin{itemize}[noitemsep,topsep=0pt,parsep=0pt,partopsep=0pt]
\item Maximum angular deviation $>$ 45$^{\circ}$.
\item Amount of overshoot $>0.5 \cdot D_{target}$.
\end{itemize}
\finalAdd{These thresholds were obtained through trial and error and can be set more or less conservatively for the designer's purposes.}

Submovements are considered \emph{interrupted} when they fall distinctly short (less than halfway) from the remaining distance to the target, or when they are classified as clutching movements.
Clutching can be determined with more or less certainty depending on the input device.
Some devices allow straightforward detection of clutching by providing touch-down and touch-up events (e.g., styluses).
Otherwise, predefined temporal thresholds can be used to detect the resetting of the end-effector, by measuring the time between two consecutive sensor events. 
In our implementation with a trackpad, we categorize a submovement as clutching when any interval of sensor events exceeds a predefined temporal threshold (130 ms).
The last submovement 
of a task was never considered as clutching.
%
%
For simplicity, submovements that are neither \emph{unaimed} nor \emph{interrupted} are deemed ``normal''.

\colorlet{LGrey}{gray!30!}
\newcommand\greycell[1]{\cellcolor{LGrey}#1}
\newcommand\nocell[0]{\greycell{No}}
%

\begin{table}[!b]
	\centering
	\small
	\begin{tabular}{|c|c|c|}
    	\hline
    	\textbf{Type of}& \textbf{Ballistic} & \textbf{Non-ballistic} \\
    	\textbf{Submovement} & ($D_{aim}=pD_{target}$) & ($D_{aim}=D_{target}$) \\
    	\hline
    	Normal & Update gains \& $p$ & Update gains only \\
		\hline
    	Interrupted & \multicolumn{2}{c|}{Update gains only} \\
		\hline
    	Unaimed & \multicolumn{2}{c|}{No update} \\
		\hline
	\end{tabular}
	\caption{
		Depending on the type of submovements, details of gain and aim point updates changes. The general objective of this classification is to exclude submovements with insufficient quality (noisy and unaimed movements) in order to facilitate the convergence of \methodname.
    }
	\label{tab:taxonomy}
\end{table}

Submovements that happen after the second normal submovement are assumed to be \emph{non-ballistic}, i.e. continuously controlled.
This assumption is based on the optimal submovement theory \cite{meyer1988optimality}, which states that in ideal cases two submovements are enough to reach to a target.
That classification is used to decide whether \methodname\ will include a submovement in future updates of the gain function and the aim point, as reported in Table \ref{tab:taxonomy}.

\finalDel{
\emph{Unaimed} submovements are excluded from the updates due to their poor aiming quality.
\emph{Non-ballistic} submovements are considered aimed directly at the target ($D_{aim}=D_{target}$) rather than at the user's implicit aim point ($D_{aim}=p\cdot D_{target}$), and therefore do not update the estimated proportion $p$.
\emph{Interrupted} submovements are treated as aiming for the estimated aim point ($D_{aim}=p\cdot D_{target}$), but falling short due to insufficient gain or poor motor planning, and therefore do not update $p$.
}

\subsection{Step 4: Speed Profiles and Aiming Errors} 
\begin{figure}
  \centering
  \includegraphics[width=1.0\columnwidth]{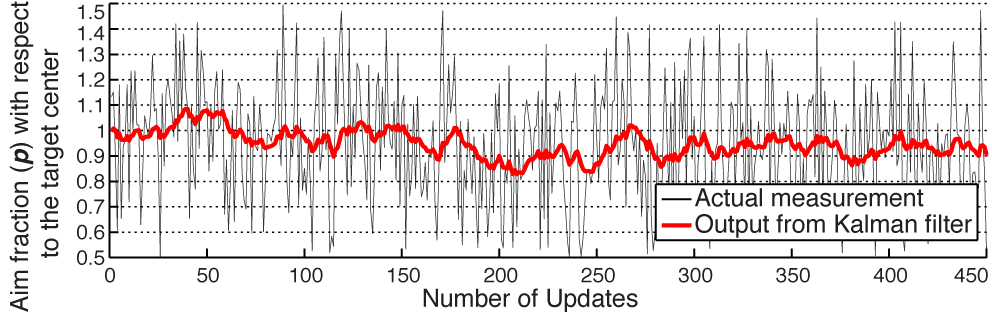}
  \caption{
  	The aim proportion $p$ is continuously updated by a Kalman filter applied after each ballistic submovement from actual measurements. The graph is showing the updates from participant 6 in Experiment 1 with \golfpart~condition (trial 1 to 60 in the first \block~).
  }
\label{fig:filtering}
\vspace{-3mm}
\end{figure}

%
\methodname\ measures the aiming error of interrupted and normal submovements.
We defined in Equation~\ref{eq:r} the aiming error $R_i$ of a submovement $i$ as the (projected) distance remaining to its estimated aim point at the end of the submovement (see Figure~\ref{fig:geometry}): $R_{i}=(D_{aim,i}-D_{c,i})$.
In non-ballistic submovements, the aim point is assumed to be the center of the target.
%
In ballistic movements, we assume that the user is aiming at a point located before the target, at a certain proportion $p$ of the remaining distance.
The true $p$ value cannot be measured directly and the $p_i$ value observed for each submovement $i$
is contaminated from high measurement noise.
Therefore, \methodname\ estimates the true $p$ value (assumed to be constant) based on the observed $p_i$ values using a Kalman filter  \cite{kalman1960new}:
%
\begin{equation}
p\sim p_i=f_\text{Kalman}\left ((D_{target,i}-D_{c,i})/D_{target,i}, p_{i-1}\right )
\label{eq:kalman}
\end{equation}
The measurement noise of the filter was set to 40, the initial process noise and $p_i$ were set to 0.2 and 1.0, respectively.
This filtering is based on the assumption of stochastic noise in human movement, whose distribution is centered at the aim point.
From the repeated observations, the filter removes stochastic noise similarly to a low pass filter (see Figure \ref{fig:filtering}).

The filter is updated after each \emph{normal+ballistic} submovement.
\finalAdd{\emph{Interrupted} submovements are treated as aiming for the estimated aim point ($D_{aim}=p\cdot D_{target}$), but falling short due to insufficient gain or poor motor planning, and therefore do not update $p$.}
%
After estimating the aim fraction $p_i$ of the current submovement, the aiming error is calculated using Equation~\ref{eq:r}:
$R_i=(p_i\cdot D_{target,i}-D_{c,i})$.

\methodname\ also records the input speeds used in every event of each submovement, in the form of a boolean array $S_i[V]$.
The range of possible input speeds is binned into $J$ speed intervals ($V_j = [v_j, v_{j+1}[, j \in [0,J-1]$) of constant width $w = v_{j+1} - v_{j}$. 
The corresponding array entry is false unless at least one event in the corresponding submovement had an instantaneous speed $v$ within that interval:
\begin{equation}
\begin{split}
    S_{i}[V_j]&= 
\begin{cases}
    1,& \text{\footnotesize if } \exists\, v: v_j \leq v < v_{j+1} \text{\footnotesize in submovement } i\\
    0,& \text{\footnotesize otherwise}
\end{cases}
\end{split}
\label{eq:selectivityrule}
\end{equation}
$R_{i}$ and $S_{i}[V_{j}]$ are used in the gain optimization process.

\subsection{Step 5: Gain Update} 
%
The principle behind \methodname, after each trial, is to increase (or decrease) the gain function when submovements in that trial undershot (or overshot) their estimated aim points, for the input speeds that were used during these submovement.
\methodname\ considers a gain function as a series of gain values associated to binned intervals of speeds input, using the same partitioning as in $S_i[V_{j}]$ (Equation~\ref{eq:selectivityrule}).
For each $j$-th speed interval, the gain function for the next trial $G_{t+1}[V_{j}]$ is updated from the gain function of the current trial $G_{t}[V_{j}]$ as follows:
\begin{equation}
G_{t+1}[V_{j}] = G_{t}[V_{j}]+\sum_{i=1}^{N_t} \Delta_{i}[V_{j}] 
\label{eq:overall}
\end{equation}
With $N_t$ the number of submovements in the trial $t$, and $\Delta_i$ the corrections calculated from each submovement's aiming error.
%
%
%
%
The amplitude of gain change $\Delta_i[V_j]$ for each speed bin is calculated using the amount of aiming error $R_i$, multiplied by a constant $C$ that defines the overall rate of gain change.
\finalAdd{\emph{Unaimed} submovements are excluded from the updates due to their poor aiming quality.}

As stated earlier, only the speed bins that have been used in the submovements of the previous pointing task will be updated ($S_i[V_{j}]$ in Equation~\ref{eq:selectivityrule}).
However, a given speed interval can be used by more than one submovement within the same pointing task.
To avoid over-favouring the speeds most commonly used, \methodname\ prioritizes submovements in descending chronological order, on the principle that submovements that occurred earlier in a trial can be sufficiently represented by their faster speed components. 
The gain change of each speed bin $V_j$ is therefore calculated in reverse order from the last submovement $N_t$, allowing only one change per bin (Equation~\ref{eq:independence}).
$\Delta_i[V_{j}]$ is calculated as expressed in Equation~\ref{eq:final}.
\begin{eqnarray}
  I_{i}[V_{j}]		&	=	&	\prod_{k=(i+1)}^{N_t} (1-S_{k}[V_{j}]) \label{eq:independence} \\
  \Delta_{i}[V_{j}]	&	=	&	C\cdot R_{i} \cdot S_{i}[V_{j}] \cdot I_{i}[V_{j}] \label{eq:final}
\end{eqnarray}
\subsection{Change Rate \textit{C}} 


In effect, the \methodname\ optimization process involves two interdependent active components: (1) gain optimization and (2) human skill acquisition. 
The system adapts the gain function to the user movements, then the user adapts his movements to the changes in the gain function, and so on.
It is therefore crucial that the updates to the gain function occur fast enough to ensure a realistically short calibration process, but slow enough to allow the user to adapt to them.
%

This is implemented in \methodname\ through the parameter $C$ in Equation~\ref{eq:final}.
If $C$ is too large, the user will not have enough time to follow up the updates in the gain function and the system might become unstable. 
If $C$ is too small, it will take too much time for the performance to converge.
Overall, the effect of $C$ on convergence speed will depend on the characteristics of the interactive system, like the scale difference between the input and output resolutions.

$C$ expresses the relationship between the amount of gain change $\delta_g$ that should occur after $M$ submovements of average aiming error $\mu_R$, for a given speed bin: 
\begin{equation}
	C=\delta_g/(M \cdot \mu_R)
	\label{eq:c}
\end{equation}
In applications of \methodname, we recommend to start with an initial $C$ value of $5\times 10^{-5}$ (in $mm^{-1}$), then increase or decrease the value \finalDel{through pilot testing based on the time pressure imposed on the optimizer and user} \finalAdd{by matching the convergence rate, observed through pilot testing, to the intended time pressure imposed on the optimizer and user}.

%
\section{Study 1: Trackpad Gain Function}

We first assess how \methodname\ fares in a laptop+trackpad setup. 
\finalDel{In particular, we were interested in if it converges to a stable function, how long it takes, and if the resulting gain function compare well to an established baseline function (Mac OS X).}
\finalAdd{We are particularly interested in whether it converges to a stable function, how long it takes, and how that resulting function compares to an established baseline (macOS function).}
%

\subsection{Participants}
We recruited 11 paid participants (4 females) aged 22 to 38 ($\mu$=27.4 years old, $\sigma$ =5.3) from the local university.
%
They were all regular MacBook trackpad users ($\mu$=4.5 years, $\sigma$=2.2) and used a trackpad 5 hours a day on average ($\sigma$=3.1).
%
Only one participant was left-handed, but typically used her right hand to control the trackpad. 
We also asked their current trackpad setting in Mac OS X (one of 10 positions on a slider, the higher the faster):
4 participants used the 4\textsuperscript{th} position (OS default), 2 used the 6\textsuperscript{th}, 4 used the 7\textsuperscript{th}, and 1 used the 8\textsuperscript{th}.
%

\subsection{Design and Dependent Variables}
The experiment followed a within-subject design with one independent variable: \emph{gain condition} = \{\refpart, \golfpart\}.
In the \refpart\ condition, we replicated each participant's everyday pointer acceleration setting (see above) using the \texttt{libpointing} \cite{casiez2011no} library.
Note that this \refpart\ condition is a strong baseline. 
It has evolved over many years of iteration and has been used for several months by participants.
In the \golfpart\ condition, the gain function was initially set to a constant gain of 1 ($G[v_t]=1~\frac{m.s^{-1}}{m.s^{-1}}$). 
This function was then updated after each trial using the \methodname~method.


Participants completed 800 trials for each \emph{gain condition}.
We counterbalanced the order of conditions across the participants.
For performance analysis, we blocked those trials into 10 \emph{block}~conditions of 80 trials with randomized Fitts's Indexes of Difficulty (ID).
We used three dependent variables to compare \refpart\ and \golfpart: \emph{trial completion time}, \emph{error rates}, and subjective assessments using the NASA-TLX \cite{hart2006nasa}.

\subsection{Task and Procedure}
The task consisted in selecting circular targets on a laptop screen using the embedded trackpad in the two different \emph{gain condition}s.
Participants were instructed to perform the tasks as quickly and as accurately as possible.
A trial ended at the first click, regardless of selection errors.
In the \golfpart\ condition, \finalDel{participants were informed about the gain adaptation.}
\finalAdd{participants were informed that changes in the gain function could occur, but not whether those changes would be positive or negative.}
\finalAdd{While this might bias participants, we thought it preferable to random reactions from (noticeable) gain function changes.}
They were also warned that pointing might feel slow or awkward in the beginning, and instructed to try to perform normally regardless. 
Participants were allowed use either tapping or pressing the touchpad to select the targets.

The target was a red disk presented on a black background.
To obtain a comprehensive picture of the observed effects, we randomized the target diameter ($[2, 11.5]$~mm), the orientation between two consecutive targets ($[0, 2\pi]$~rad), and the ID of each task ($[2, 5.5]$~bits).
%
%
%
We used the following randomization process after each successful click $(x,y)$:
\begin{center}
\begin{tabular}{r@{ : \hspace{5mm}}l@{\hspace{1cm}}}
	\hline
	1	&	$I\!D \leftarrow$ random(2, 5.5);\\
    2	&	\textbf{do:}\\
    3	&	\hspace{5mm} $W_c \leftarrow$ random(2, 11.5);\\
    4	&	\hspace{5mm} $x_c \leftarrow$ random(0, screen.width);\\
    5	&	\hspace{5mm} $y_c \leftarrow$ random(0, screen.height);\\
    6	&	\hspace{5mm} $I\!D_c \leftarrow$ $\log_2\left(1+\frac{\text{dist}[(x, y), (x_c, y_c)]}{W_c}\right)$;\\
    7	&	\textbf{until} $| I\!D_c - I\!D | < 0.1$\\
    \hline
\end{tabular}
\end{center}
This ensured uniform distributions of IDs and orientations, providing comparable performance datasets.


Participants sat on a regular office chair that they could adjust, and used a laptop placed on a desk.
They first filled a preliminary questionnaire about their regular MacBook trackpad usage and proceeded to the task.
Participants were instructed to take a break every 80 selections and answered a NASA-TLX form about their performance since the last break.
The experiment lasted about one hour per participant.

\subsection{Apparatus}
We ran the experiment on a MacBook Pro laptop (2012 version) running Mac OS X 10.11 with a integrated trackpad.
The size of the display was 35.8 cm $\times$ 24.7 cm (1280 $\times$ 800 pixels).
The experiment was coded in C++.
The optimization speed parameter $C$ in Eq.~(\ref{eq:overall}-\ref{eq:final}) was set to $6.4 \times 10^{-5}$~mm$^{-1}$ after pilot tests.
The \refpart\ functions were obtained from \texttt{libpointing}~\cite{casiez2011no}, which was also used for cursor coordinate calculations in this \emph{gain condition}.
We used \texttt{libpointing}'s `subpixel' option that transfers the remainder of the last calculated (floating) cursor coordinates to the next time step.
We implemented the same subpixel mechanism in \methodname, 
and discretized the range of input speeds into bins of 0.0079 m/s as \refpart~condition.
We set the processing noise parameter of the Kalman filter at 0.2, and the sensor noise parameter at 40.0.
The refresh rate of the display was 60 fps.

\subsection{Results}

\subsubsection{Data Processing}
We excluded one participant from our dataset, who displayed irregular initial performance in both \emph{gain condition}s as well as inconsistent aiming behavior (high variance in parameter $p$, Eq.~(\ref{eq:p})).
%
%
Her ballistic movements were also notably slower:
while the other participants' input was 96\% higher in the first \block\ of \golfpart\ than of \refpart\ (due to the lower initial gain function),
hers was only 15\% higher in \golfpart\ for that block.
As a result, while \methodname\ did improve her performance over time, the improvement was markedly slower.
We exclude that participant from the following analyses (N=10), and take note that movement consistency can play an important role in the improvement rate of \methodname.
%
%

We analyzed 8,000 trials per \emph{gain condition}, corresponding to 19,766 submovements in the \refpart\ condition (2.47 per trial) and 19,828 submovements in the \golfpart\ condition (2.48 per trial).
In the \refpart\ condition, 4,349 submovements (22.0\% of all submovements, 0.54 times per trial on average) were evaluated as \emph{interrupted},  4,774 submovements (24.1\% of all submovements, 0.60 times per trial) were evaluated as \emph{unaimed} submovements, and 6,031 submovements (30.5\% of all submovements, 0.75 times per trial) were evaluated as \emph{non-ballistic}.
%
In the \golfpart\ condition, 4,275 submovements (21.6\%, 0.53 times per trial) were \emph{interrupted} submovements, 3,577 (18.0 \%) were evaluated as \emph{unaimed} submovements, and 6,769 submovements (34.1\% of all submovements, 0.75 times per trial) were evaluated as \emph{non-ballistic}.
The overall average of aim point ($p$) estimated from Kalman filter was 0.94 ($\sigma$ =0.033) for \golfpart~condition.
The completion time and error rates are averaged by blocks of 80 trials before analysis.
The error rate is defined as the percentage of trials that misselected the target among all trials.

We used two-way repeated-measure ANOVAs for the following analysis.
Although each participant used their personal gain setting in \refpart\ condition, the gain setting had no significant effect on trial completion time (F(1,7)=0.686, p=0.435).
The order of gain conditions also had no significant effect on overall trial completion time (F(1,7)=0.045, p=0.838).
So we exclude the above factors from the analysis.

\subsubsection{Error Rates}
The mean error rate was 9.3\% ($\sigma$=2.2) for \refpart\ and 9.3\% ($\sigma$=2.8) for \golfpart.
We attribute that overall high error rate to the higher number of small targets generated by the target randomization scheme, as has been reported in previous studies \cite{wobbrock2008error,zhai2004speed}.
We found no significant effect of \emph{gain condition} (F(1,9)=0.001, p=0.984) and \emph{block} (F(1,9)=1.261, p=0.29) on error rate.
We also found no interaction effect between \emph{block} and \emph{gain condition} on error rate (F(9,81)=1.02, p=0.432). 
We conclude that the participants maintained comparable accuracy throughout the experiment.

\subsubsection{Trial Completion Time}
\begin{figure}[h]
  \centering
  \includegraphics[width=0.95\columnwidth]{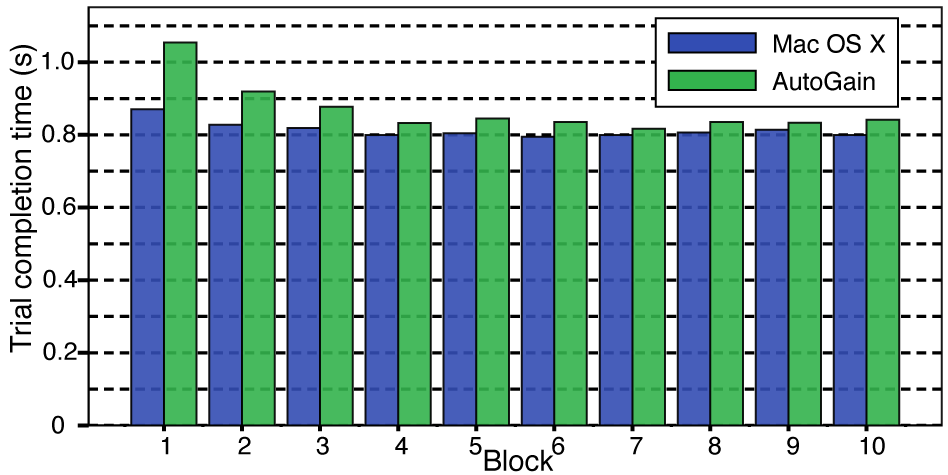}
  \caption{Study 1: The average difference in trial completion time between \golfpart~and \refpart~was 56~ms. The difference became non-significant from \block~7 (p=0.35) onward.}
  \label{fig:S1trialcompletiontime}
\end{figure}

We found a significant effect of of \emph{gain condition} on trial completion time (F(1,9)=40.68, p$<$0.001).
The average completion time was 813.5~ms ($\sigma$=30.7) for \refpart\ and 869.1~ms ($\sigma$=26.4) for \golfpart.
This result is expected: in the \golfpart\ condition, participants started from a slow gain of 1 $\frac{m.s^{-1}}{m.s^{-1}}$.
However, the difference was only about 50 ms. 
This is on a similar level with a difference reported between OS X function and Windows in a previous study ($=50$ ms) \cite{casiez2011no}. Also, it was quickly improved after few blocks (see Figure \ref{fig:S1trialcompletiontime}).

The interaction effect between \emph{block} and \emph{gain condition} on trial completion time was significant (F(9,81)=10.24, p$<$0.001).
Pairwise comparison showed that the differences in trial completion time between \emph{gain condition}s become not significant from \block~7 onward (p=0.35, see Figure \ref{fig:S1trialcompletiontime}).
\finalAdd{The converged trial completion time of \methodname~after \block~7 remained slightly higher than the baseline by 26.9 ms on average (n.s.).}
Completion times after \block~7 were as follows:
\begin{table}[h]
	\centering
	\footnotesize
    \begin{tabular}{c|ccc}
        Block \#& \refpart\ (ms) & \golfpart\ (ms) & p \\
        \hline
        7		& 799.35 ($\sigma$=34.8) & 816.95 ($\sigma$=26.4) & .35 \\
        8		& 806.44 ($\sigma$=28.8) & 835.34 ($\sigma$=27.8) & .065 \\
        9		& 814.28 ($\sigma$=38.9) & 833.54 ($\sigma$=32.4) & .141 \\
        10		& 799.32 ($\sigma$=30.2) & 841.31 ($\sigma$=32.4) & .137
    \end{tabular}
\end{table}


\subsubsection{Workload Metrics}
\begin{figure}[h]
\centering
\includegraphics[width=1.0\columnwidth]{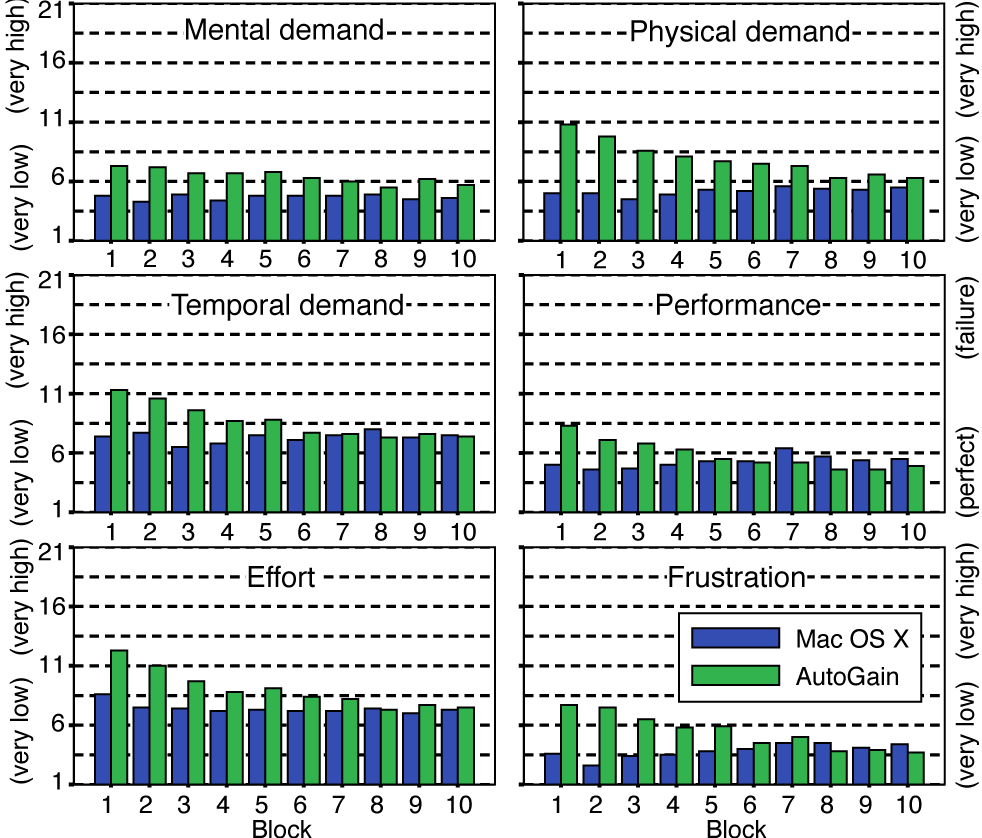}
\caption{The difference in subjective ratings between \refpart~and \golfpart~becomes insignificant after 6 to 7 blocks in the trial. Users reported performance being higher in \golfpart~ than in \refpart~after \block~7.}
\label{fig:subjectiveratings}
~\vspace{-3mm}
\end{figure}

For simplicity we report Raw TLX values \cite{hart2006nasa} (Figure \ref{fig:subjectiveratings}).
We found a significant effect of \emph{gain condition}s on mental demand (F(1,9)=6.40, p=0.032), physical demand (F(1,9)=21.69, p=0.001), and effort (F(1,9)=9.30, p=0.014).
However, from the interaction effect between \emph{block} and \emph{gain condition}, these differences pertained to the beginning half of the study and became non-significant in the later half. 
(Table \ref{tab:tlx}).
Differences were not significant for temporal demand (F(1,9)=2.97, p=0.12), performance (F(1,9)=1.04, p=0.37), and frustration (F(1,9)=4.47, p=0.064).
%
\begin{table}[h]
\footnotesize
\begin{center}
    \begin{tabular}{ c | c c c}
    Measure&\refpart&\golfpart&Insignificant from\\ 
    \hline
    Mental demand&4.7 ($\sigma$=0.9)&6.4 ($\sigma$=1.3)&\block~6 (p=.062)\\ 
    Physical demand&5.2 ($\sigma$=0.8)&7.9 ($\sigma$=1.2)&\block~7 (p=.063)\\ 
    Effort&7.4 ($\sigma$=1.3)&9.0 ($\sigma$=1.4)&\block~6 (p=.133)\\ 
    \end{tabular}
    \caption{Mean and standard deviation for overall TLX measure in both \emph{gain condition}s (lower is better). In every measure, the difference became insignificant after a certain \block.}
    \label{tab:tlx}
    \end{center}
\end{table}

\subsection{Summary}

\begin{figure}[h]
	\centering
	\includegraphics[width=1.0\columnwidth]{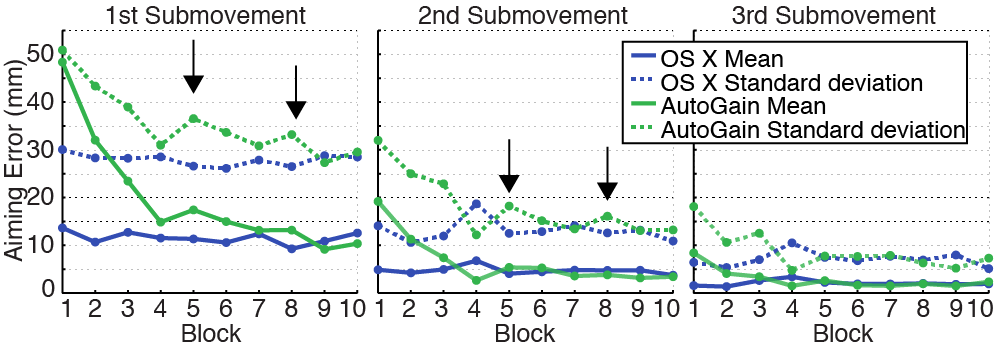}
	\caption{
  		\methodname~gradually reduced aiming error in submovements to levels similar to the \refpart\ (Mac OS X). 
        Fluctuating dynamics in aiming error were observed in \block s 5 and 8 (black arrows).
	}
	\label{fig:distanceresidual}
\end{figure}

In 30 minutes of intensive use, \methodname\ produced a well-performing gain function starting from scratch. User performance and workload ratings were comparable to that of Mac OS X trackpad function, which has been hand-tuned since 1994 (1st MacBook trackpad) and \finalAdd{replicated the participants' daily settings.}
The data lend evidence to the main goal of \methodname\, 
which is to reduce aiming errors as a means to improve pointing performance.
Figure \ref{fig:distanceresidual} shows that these errors are indeed being reduced over time. 
Performance between \methodname\ and the baseline became similar at the same time as aiming errors, around \block~7.
%
%

\begin{figure}[h]
	\centering
	\includegraphics[width=0.95\columnwidth]{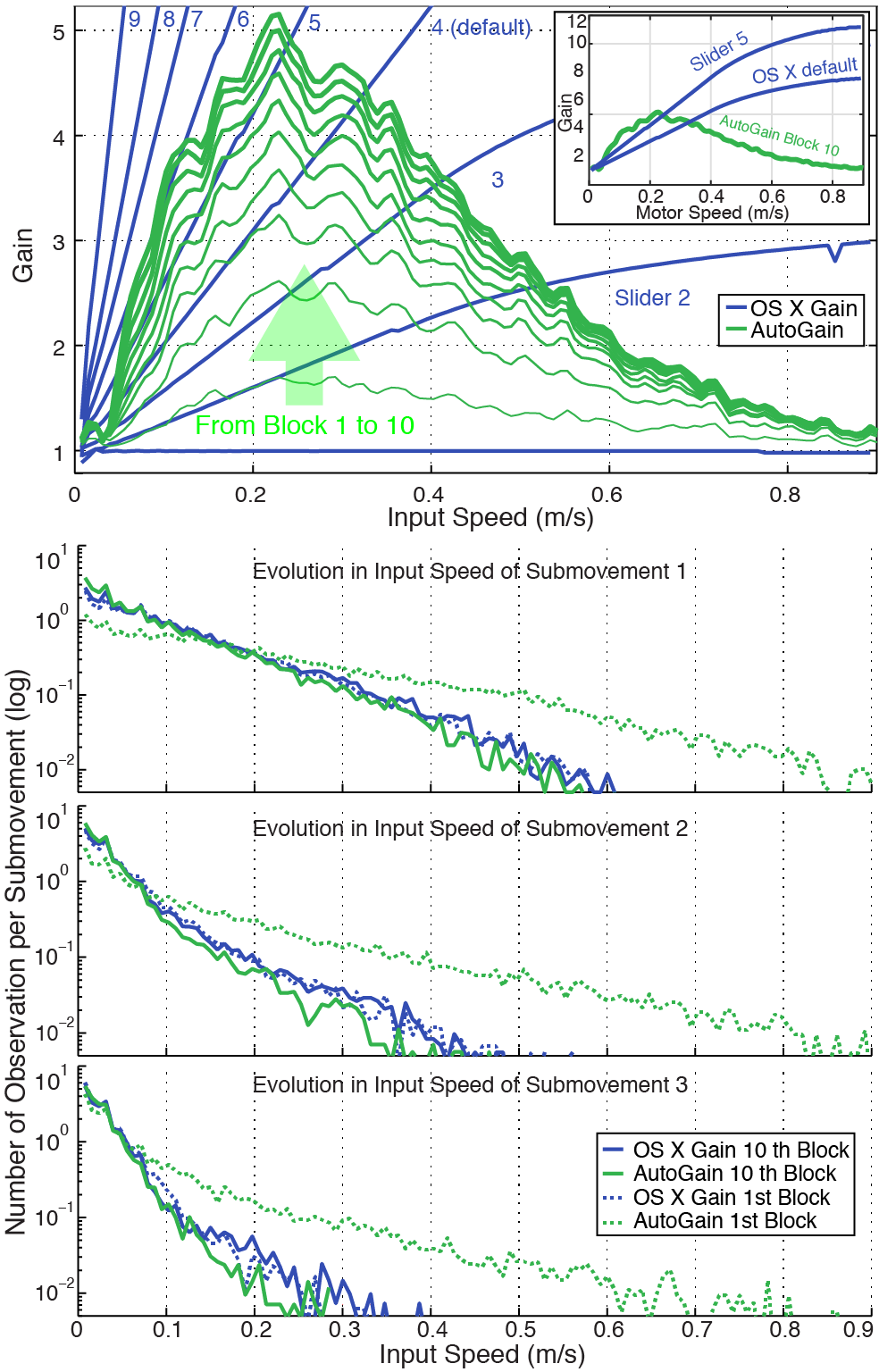}
	\caption{
		Top: \methodname~gradually updates the gain function until convergence (no more improvements in aiming error). 
		Bottom: the evolution in input speed averaged for all participants. After 10 blocks of trials, participants were utilizing the same speed intervals with \methodname\ as with the OS X function.
	}
	\label{fig:gainfunction}
\end{figure}

Figure \ref{fig:gainfunction} shows the gain functions obtained by \golfpart\ alongside the Mac OS X functions used in \refpart\ (top),
%
%
%
and illustrates how the input speeds used in \golfpart\ and \refpart\ became similar by the end of the experiment (bottom), especially between 0 and 0.3 m/s.
This range also corresponds to the regions where the functions of \golfpart\ and \refpart\ were most similar, both in value and slope.

The much steeper downward slope of the \methodname-produced function after 0.3 m/s reflects the smaller number of input events featuring these input speeds:
\methodname\ updates the gain corresponding to an input speed interval only when this interval was used in the previous pointing task. 
However, the comparable completion times obtained between the two \emph{gain condition}s tend to indicate that the gains above commonly used speeds is of lesser importance for overall performance.

\section{Study 2: Gain Function For An Emerging Device}

The objective of this study is to assess \methodname's ability to produce usable gain functions on input devices 
that were not primarily designed for cursor control---in this case, a Leap Motion controller.
Indirect pointing with this device has already been explored, e.g. in \cite{liu15}, using a hand-tuned sigmoid gain function based on \cite{nancel15}.
We are interested to see if \methodname\ stabilizes to a gain function with satisfying performance and user feedback, and how that resulting function fares compared to the state-of-the-art function for this device.

\subsection{Participants}

We recruited 10 paid participants (3 females and 7 males; 23 years old in average, $\sigma$=3.33) from a local university.
None of them had experience with Leap Motion. 
One used his left hand to control the cursor. 
Four had corrected-to-normal vision.

\subsection{Design}

This experiment design is similar to Study 1: it followed a within-subject design with one independent variable, \emph{gain condition}: \golfpart~and \sigpart.
We counterbalanced the order of conditions across the participants.
Each participants was to perform 800 trials for each gain condition, which we blocked into 10 \emph{block}s of 80 trials.

The initial gain function for \golfpart\ was a one-to-one constant ($G(v)=1~\frac{m.s^{-1}}{m.s^{-1}}$) between the speeds of the index finger and of the cursor.
The \sigpart\ condition used the function 
$G(v)=G_{min}+(G_{max}-G_{min})/(1+e^{-\lambda(v-V_{inf})})$
with parameters tuned based on the recommendations in \cite{nancel15}.

\subsection{Task and procedure}

The task was the identical to Study 1.
Participants used their dominant hand over a Leap Motion tracker to move the cursor, and placed their non-dominant hand over the built-in keyboard to click and clutch.
Clicking was performed by pressing the space bar.
The cursor only moved when the C key was held pressed; clutching was performed by releasing that key, relocating the finger, and resuming pointing by pressing C again.
The finger movements were interpreted as parallel to the display, i.e. on a vertical plane.
This ensured forward/up visual compatibility \cite{phillips2005forward} between hand and cursor movements. 

\begin{figure}[h]
  \centering
  \includegraphics[width=1.0\columnwidth]{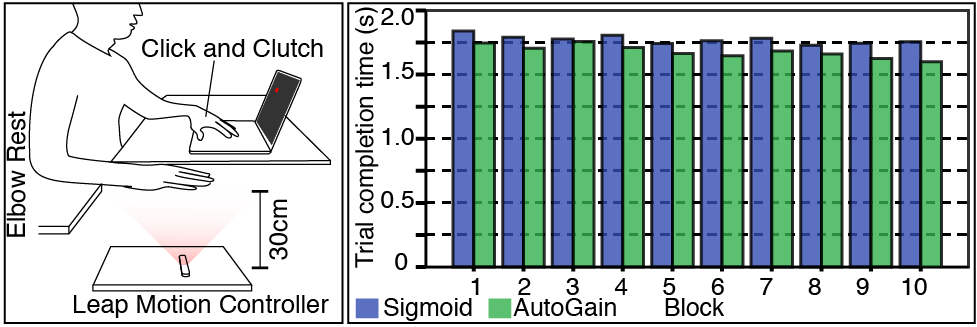}
  \caption{
  	(left) Apparatus and trial completion times in Study 2. (right) Completion times per Condition and Block.
  }
  \label{fig:timeleap}
\end{figure}

During the tasks, participants rested their dominant elbow on a stand aside the desk (at desk height) to reduce fatigue, with the Leap Motion device facing up below their hand and 30 cm below desk height (see Figure~\ref{fig:timeleap}-left).
%

For each participant, we first calibrated the Leap Motion tracker using its inbuilt calibration software before starting the experiment.
An experiment assistant then briefly demonstrated the pointing mechanism.
%
Participants sat on a regular office chair that they could adjust, and used a laptop placed on a desk. 
They first filled a preliminary questionnaire asking about their previous experience with mid-air pointing.

Participants were instructed to take 10 minutes of practice session with an initial constant gain of 1 before the main part of the experiment started. 
The gain function was not updated during the practice session.
During the main session, participants were invited to take breaks every 80 selection tasks.
The experiment lasted about 90 minutes per participant.


\subsection{Apparatus}

The experiment was coded in C++ and run on the same computer as in Study 1.
The cursor was controlled using mid-air hand movements tracked by a Leap Motion controller, (software version 2.3.1+31549).
We used the Leap Motion in ``Robust tracking mode'' and smoothened its raw input using the \texttt{1~\euro~filter} \cite{casiez20121}.
Following Casiez \etal{}'s tuning guidelines, we used $10^{-5}$ for minimum cutoff frequency and 0.05 for beta parameter.

For \golfpart, the change rate parameter $C$ was set to 3.6$\times10^{-5}$~mm$^{-1}$ after pilot tests; 
this is lower than Study 1 to account for the possible harder learning and increased positional noise, compared to traditional trackpad input.
We again set the processing noise parameter of the Kalman filter at 0.2, and the sensor noise parameter as 40.0.
The refresh rate of the display was 60 fps.

For dynamic tracking of moving hands, the spatial resolution of the Leap Motion controller was reported at around 0.7 mm \cite{weichert2013analysis}.
Therefore, to avoid being sensitive to noise, we discretized the range of input speeds into bins of 0.06 m/s (1 mm/count, considering Leap Motion's 60Hz frequency).

For \sigpart, we could not reuse the function parameters from \cite{liu15} as their use-case was significantly different: participants were standing, the task was performed on a large display, participants had less training (36 trials), clicking and clutching were triggered by hand gestures with inconsistent recognition accuracy, etc.
We adapted the tuning recommendations from \cite{nancel15} and obtained the following parameters: $G_{min}= 0.48$, $G_{max}=2.93$, $V_{inf}= 0.219$, $\lambda=12.54$.
\todo{Are Gs expressed in mm-on-screen per mm-in-real-life?}

\subsection{Results}

\subsubsection{Data Processing}

We gathered 34,193 submovements (4.27 times per trial) in the \sigpart~condition.
Among them, 2,829 were \emph{interrupted} (8.27\% of all submovements, 0.35 times per trial), 14,657 were \emph{unaimed} (42.9\%, 1.83 times per trial), and 14,668 were \emph{non-ballistic} (42.9\%, 2.44 times per trial).
We gathered 31,969 submovements (4.00 times per trial) in \golfpart.
Among them, 2,440 were \emph{interrupted} (7.63\% of all submovements, 0.31 times per trial), 14,226 were \emph{unaimed} (44.5\%, 1.78 times per trial), and 14,010 were \emph{non-ballistic} (43.8\%, 2.24 times per trial).
The overall average of aim point ($p$) estimated from Kalman filter was 1.02 ($\sigma$= 0.021). In the following analysis, we used ANOVA and applied Greenhouse-Geisser correction when the assumption of sphericity was violated.

%

\subsubsection{Error Rates}

There were no significant effect of \emph{gain condition} (p=0.063, F(1,9)=4.495) or \emph{block} (p=0.559, F(3.23,29.07)=0.717) on error rate.
%
\finalAdd{Participants in \golfpart~had slightly higher error rates ($\mu$=7.71\%, $\sigma$=6.7\%) than in \sigpart~condition ($\mu$=5.09\%, $\sigma$=4.4\%, not significant).}
\finalDel{The mean was 5.09\% ($\sigma$=4.4\%) in the \sigpart and 7.71\% ($\sigma$=6.7\%) in the \golfpart~condition.}

\subsubsection{Trial Completion Time}
In Experiment 2, unlike Experiment 1, both gain functions were unfamiliar to the user. Therefore, we first eliminated the learning effect seen in the block and compared the two gain functions.
There was a significant effect of \emph{block} (p$<$0.001, F(9,81)=5.033).
Overall average of trial completion time was 1.77 s ($\sigma$=0.152 s) in the \sigpart~condition and 1.68 s ($\sigma$=0.136 s) in the \golfpart~condition.

To test the effect of \emph{block} on trial completion time, we used Helmert contrast 
\finalAdd{, which consists of comparing the mean of each level of a factor (except the last) to the mean of subsequent levels.}
The effect of \emph{block} on trial completion time becomes insignificant from block 8 (p=0.670).
Hence we averaged block 8, 9, and 10 in the following analysis.
%

For averaged blocks 8, 9, and 10, the effect of \emph{gain condition} on trial completion time was significant (p=0.008, F(1,9)=11.469).
The mean trial completion time in the \sigpart~condition was 1.74 s ($\sigma$=0.144 s), and 1.62 s ($\sigma$=0.1 s) in the \golfpart~condition.
Means and standard deviations for each input condition are summarized in the table below, for blocks 8, 9, 10:
\begin{table}[h]
	\centering
	\footnotesize
	\begin{tabular}{c|ccc}
		Block \#& \sigpart\ (ms) & \golfpart\ (ms)\\
		\hline
		8		& 1782.72 ($\sigma$=137.4) & 1660.01 ($\sigma$=120.2)\\
		9		& 1744.58 ($\sigma$=171.3) & 1625.19 ($\sigma$=115.8)\\
		10		& 1756.56 ($\sigma$=135.8) & 1599.06 ($\sigma$=85.0)
    \end{tabular}
\end{table}
We also analyzed the error rate again after averaging blocks 8, 9, and 10.
The effect of \emph{gain condition} on error rate remained not significant (p=0.122, F(1,9)=2.915).
The error rate in the \sigpart~condition for averaged blocks 8, 9, and 10 was 5\% ($\sigma$= 4.9\%), and 7.4\% ($\sigma$=7.8\%) in the \golfpart~condition.

\subsection{Discussion}


While there is no reference gain function for indirect pointing with Leap Motion, we compared \methodname\ to a sigmoid function tuned using recommendations from previous work~\cite{nancel15}.
%
\methodname\ was able to offer significant improvements, about 6.8\%, in completion time when compared to the sigmoid function, without a significant decrease in accuracy.
\finalAdd{We note that the performance in \sigpart~conditions started already lower than \methodname~from the first block, which suggests a shortcoming of hand-tuning methods, that ultimately rely on subjective impressions rather than objective criteria.}

\begin{figure}[!t]
  \centering
  \includegraphics[width=0.95\columnwidth]{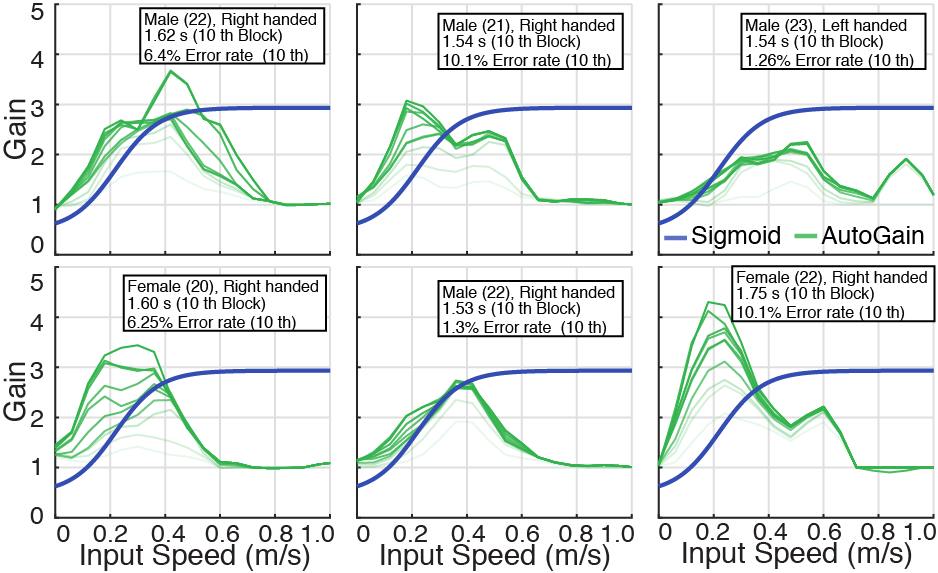}
  \caption{\methodname\ adapts to each individual user. One left handed user converged to a very unique gain function, whose trial completion time was lower than other participants.  
  \mn{Please add 95\% confidence intervals.}}
  \label{fig:trialcompletiontime}
  ~\vspace{-6mm}
\end{figure}

As in the first study, the gain functions produced by \methodname\ displayed significant completion time improvement from the initial block (here up to 20.2\%), and reasonable completion times (here around 1.6 second).
Figure \ref{fig:trialcompletiontime} shows the evolution of the gain functions that it produced for a subset of our participants after each block.
Each function appears to converge to a particular shape, which might reflect the movement specificities and pointing strategy of individual participants.

\section{Study 3: Field Study with a Windows App}


To assess whether \methodname\ can improve long-term pointing performance in realistic setups, we ran a one-month longitudinal experiment with two participants, in which it gradually updated the gain function of a desktop OS (Windows 10), starting from each participant's pre-existing gain function.
We developed an application (available as an open source project at \url{https://github.com/SunjunKim/AutoGain}) that bypasses the system's gain function and replaces it with a custom one.

\begin{figure}[!t]
\centering
\includegraphics[width=1.0\columnwidth]{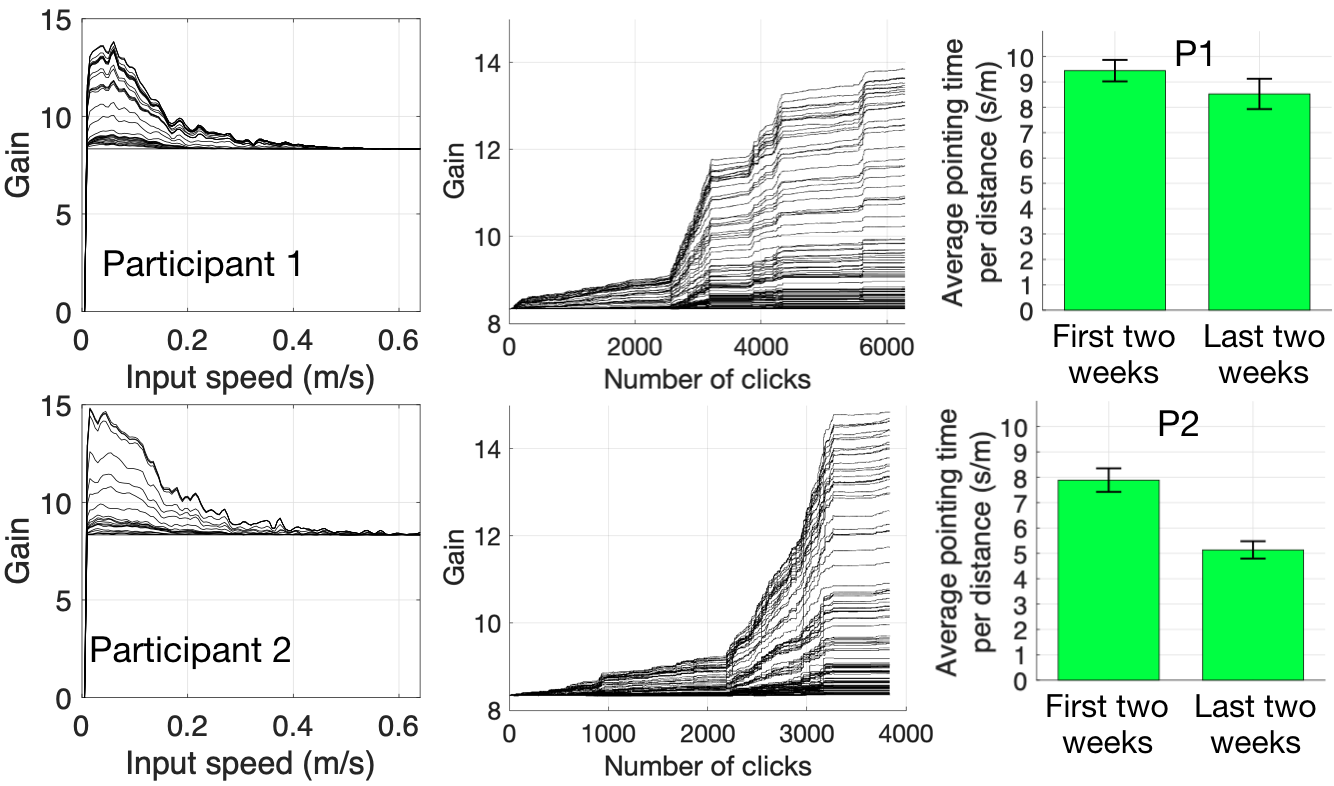}
\caption{
    \methodname~deployed as a Windows application for everyday use.
    Left: evolution of the gain functions (upward), one line per 80 clicks.
    Middle: evolution of each individual gain bin.
    Right: Evolution of pointing performance, in s/m to account for uncontrolled tasks.
    Over a month, participants' pointing time per meter improved by an average of 21.4\% by \methodname. 
    Error bars represent 95\% confidence interval.
    }
\label{fig:application}
\end{figure}

\subsection{Method}
We conducted a 1-month study with two volunteers (1 female, 1 male, ages 27 and 28).
We installed the application on their office desktop computers, both operated by mice.
Participants were informed that the behaviour of their cursor behavior might change throughout the study.
We gave them no special instructions besides using their computer normally for about a month.
The gain function they both usually used was the Windows default function, with the acceleration turned off.
%
The optimization speed parameter $C$ (Eq.~\ref{eq:c}) was set to $5 \times 10^{-5}$~mm$^{-1}$, similar to the previous two user studies.

In everyday situations it is difficult to find out where the center of the target the user clicked was and when the first submovement of the aiming movement for that target began.
Several methods have been proposed (see e.g. \cite{fittswild, evans2012taming}), but not formally compared.
For simplicity, and because \methodname\ does not require precise information about target shape or size, we applied two heuristics: (1) the clicked position approximates the target center, and (2) the submovement with the highest speed peak approximates the start of the pointing motion \cite{evans2012taming}.

\subsection{Results}


6,285 (P1) and 3,839 (P2)
click events were observed in the collection period.
\finalAdd{Unlike previous studies, participants barely noticed changes in mouse pointer movement. This is because clicks, and therefore gain function updates, occurred at much lower rates than during a controlled experiment. Also, being an "in-the-wild" study, the percentage of unaimed submovements (30.4 \%) that did not lead to an update of the gain function was higher than in Study 1 with the trackpad (18 \%).}

The gain functions in each computer evolved in a similar fashion (see Fig.~\ref{fig:application}), with a steep increase around clicks \#2500-3500 surrounded by gentler slopes.
%
We hypothesize that participants first got used to slightly higher gains, progressively learning to take advantage of it by using smaller input speeds.
This prompted \methodname\ to rapidly increase gains up to a point where it no longer measured an aiming advantage.
The same process repeated afterwards, with less drastic increases, as users continued to adapt their movements.

%

To assess the evolution of pointing performance, we compared overall pointing times between the first and last two weeks of the experiment.
Since task difficulty can vary greatly, and we cannot reliably assess target size from pointing movement only, we considered pointing time as a function of pointing distance.
The two participants took resp. 9.45 ($\sigma$=12.13) and 7.89~s/m ($\sigma$=10.72) for the first two weeks, but 8.53 ($\sigma$=17.15) and 5.14 ($\sigma$=7.91)~s/m for the last two, i.e. improved by resp. 9.74 \% and 34.85 \%.
Participants reported no difficulty using the mouse while \methodname~was on.

\section{Summary and Limitations}

We introduced a method inspired by motor control theories to automatically optimize a gain function for a  user, using only pointer motion as input and making minimal assumptions about the shape of that function. 
\methodname\ gradually adapts the gain function to minimize aiming error at submovement level. 
%
\finalDel{It can produce efficient and stable gain functions in less than an hour of use with, also with novel input systems that do not have yet a well-established reference function.}
In Study 1, \methodname\ produced gain functions for trackpads yielding performance comparable to that of widely used commercial gain functions, from scratch, 
in about 30 minutes of active use.
In Study 2, it produced gain functions for a Leap Motion controller, for which no well-established reference function yet exists, and yielding faster task completion than a sigmoid function from the literature, hand-tuned to the best of our abilities.
\methodname\ produced notably distinct gain functions for different participants, indicating potential to adapt to individual movements and pointing strategies.
Study 3 showed that the obtained functions converge in real use and improve pointing performance over users' default function.
%

\methodname\ opens up exciting possibilities for personalized pointing facilitation techniques, in both existing and new systems.
It could be used to accelerate the adoption of novel input devices for which no reference gain function yet exists.
Pointing facilitation techniques could adapt independently in different context, e.g. for games or drawing apps.
To this end, we publish \methodname's Windows application as open source.
In future work, we plan to apply \methodname\ to more varied platforms, such as large displays controlled in mid-air, or high-precision machinery.
Exploring a broad range of such systems might help us refine its initialization steps, e.g. the rate at which gain updates occur ($C$ in Eq.~\ref{eq:final}), which would both accelerate the convergence of the optimization process and, possibly, the efficiency of the resulting gain functions.

Finally,
while \methodname\ was able to show statistically significant improvements over a baseline (Study 2), 
more empirical work is needed to disentangle the effects of gain adaptation from users' learning, both of which affect each other.

\finalAdd{
\section{Acknowledgements}
This research was funded by the National Research Foundation of Korea (2017R1C1B2002101), Korea Creative Content Agency (R2019020010), and the European Research Council (ERC) under the European Union's Horizon 2020 research and innovation programme (No 637991)}

\balance{}

\bibliographystyle{SIGCHI-Reference-Format}
\bibliography{proceedings}

\end{document}